\documentclass[pre,showpacs,twocolumn,superscriptaddress,floatfix]{revtex4-1}
\usepackage{amssymb,amsfonts}
\usepackage{amsmath}
\usepackage[pdftex]{graphicx}
\usepackage[T1]{fontenc}
\usepackage{lmodern}
\usepackage{color}
\usepackage{refstyle}

\begin{document}
\title{Cluster virial expansion and electron-hydrogen molecule scattering}
 \author{ Y. A. Omarbakiyeva}
 \email{yultuz.omarbakiyeva@uni-rostock.de}
  \affiliation{Institute of Physics, University of Rostock, D-18051, Rostock, Germany}
 \affiliation{International IT University, 050040, Almaty, Kazakhstan}
 \author{H. Reinholz}
\affiliation{Institute of Physics, University of Rostock, D-18051, Rostock, Germany}
\author{G. R\"opke}
\affiliation{Institute of Physics, University of Rostock, D-18051, Rostock, Germany}
\date{\today}

\begin{abstract}
The equation of state of partially ionized hydrogen plasma is considered with special focus on
the contribution of $e-\rm H_2$ interaction. Within a cluster virial expansion, the Beth-Uhlenbeck formula
is applied to infer the contribution of bound and scattering states to the temperature dependent
second virial coefficient. The scattering states are calculated using the phase expansion method with
the polarization interaction model that incorporates experimental data for the $e-\rm H_2$ scattering cross section. 
We present results for the scattering phase shifts, differential
scattering cross sections, the second virial coefficient due to $e-\rm H_2$ interaction. The influence of this
interaction on the composition of the partially ionized hydrogen plasma is confined to the
parameter range where both the $\rm H_2$ and the free electron components are abundant.
\end{abstract}
\maketitle

\section{Introduction\label{sec:Introduction}}
The equation of state describing fundamental characteristics of matter attracts significant attention of researchers from multiple disciplines. 
For instance, the calculation of thermodynamic properties of plasmas in the warm dense matter region is necessary to solve problems of high energy density physics 
\cite{drake} and to model the planetary and stellar interiors \cite{lewis,redmer}.  Correlations and bound state formation are of relevance since simple treatment of 
plasma using perturbation methods is not possible in this region. 
 In particular, non-ideal contributions are involved due to the interaction between charged and neutral particles, 
which can be treated either within the chemical or the physical picture. 

In the chemical picture, the plasma is assumed to consist of well-defined, reacting particles - electrons, ions, atoms and molecules. Interactions are described via 
effective short-range potentials for neutral particles and long range potential between charged particles. 
The thermodynamical characteristics of non-ideal plasma can be represented by the free energy, 
which is calculated on the basis of different pseudopotential models for certain pair interaction.   
Usually, the non-ideal part of the free energy consists of the contributions 
for Coulomb interaction (electron-electron, ion-ion, ion-electron), 
polarization interactions between charged and neutral particles and short-range interactions between neutrals.
This model has been successfully applied 
\cite{HRGFI} to investigate  properties of partially ionized plasmas. Nevertheless 
it should be 
systematically studied within quantum-statistical methods to avoid inconsistencies such as double counting effects.

In the physical model, the fundamental structural elements are the electrons and protons with Coulomb interaction, 
and the composite particles, atoms, molecules and other heavier components are obtained from few-body wave equations. The latter are assumed 
to consist of fundamental particles and their properties should be determined by solving the corresponding Schr\"odinger equation.  
Within the physical picture, virial expansions (with respect to density or fugacity) 
can be evaluated.  
In the density virial expansion, the second virial coefficient is determined by pair interactions.
Interactions of electrons with the neutral composite 
particles appear in higher orders (third virial coefficient etc.). Alternatively,  
the contribution of neutral particles can be taken into account within a cluster virial expansion. 
In the fugacity expansion,  formation of bound
states (clusters) are consistently included. For instance, in the low density limit two-particle bound states are stable. 
Therefore, it is possible to consider the bound states as
new particles. We switch from the physical picture to the chemical picture what means the partial summation of 
ladder diagrams that describes the formation of bound states with a Green function approach. 
The three particle interaction in the physical picture will be considered as effective two 
particle interaction in the chemical picture after inclusion of cluster states. 

The cluster virial expansion has been described in details in a previous paper \cite{OFRR}. 
The electron-atom interaction was studied from a microscopic point of
view. Different pseudopotentials were compared and empirical data for separable potentials were given. 
 With the help of the Beth-Uhlenbeck formula \cite{BU} it has been shown that the
second virial coefficient in the electron-atom channel is related to scattering phase shifts  as well as
bound states. In contrast to previous approaches, results for
the second virial coefficient in the $e - \rm H$ channel are not based on any pseudopotential
models but are directly derived from measured scattering data. Simultaneously, the contribution of the bound state $\rm H^{-}$ 
was included. The use of experimental data as an input for the Beth-Uhlenbeck
formula avoids any empirical parameters and may be considered as low-density benchmark for any equation of state. In this present work, 
this approach will be extended to include further components of the plasma, in particular $\rm H_2$ molecules interacting with electrons.

\begin{figure*}[htp]
\includegraphics[width=0.65\textwidth]{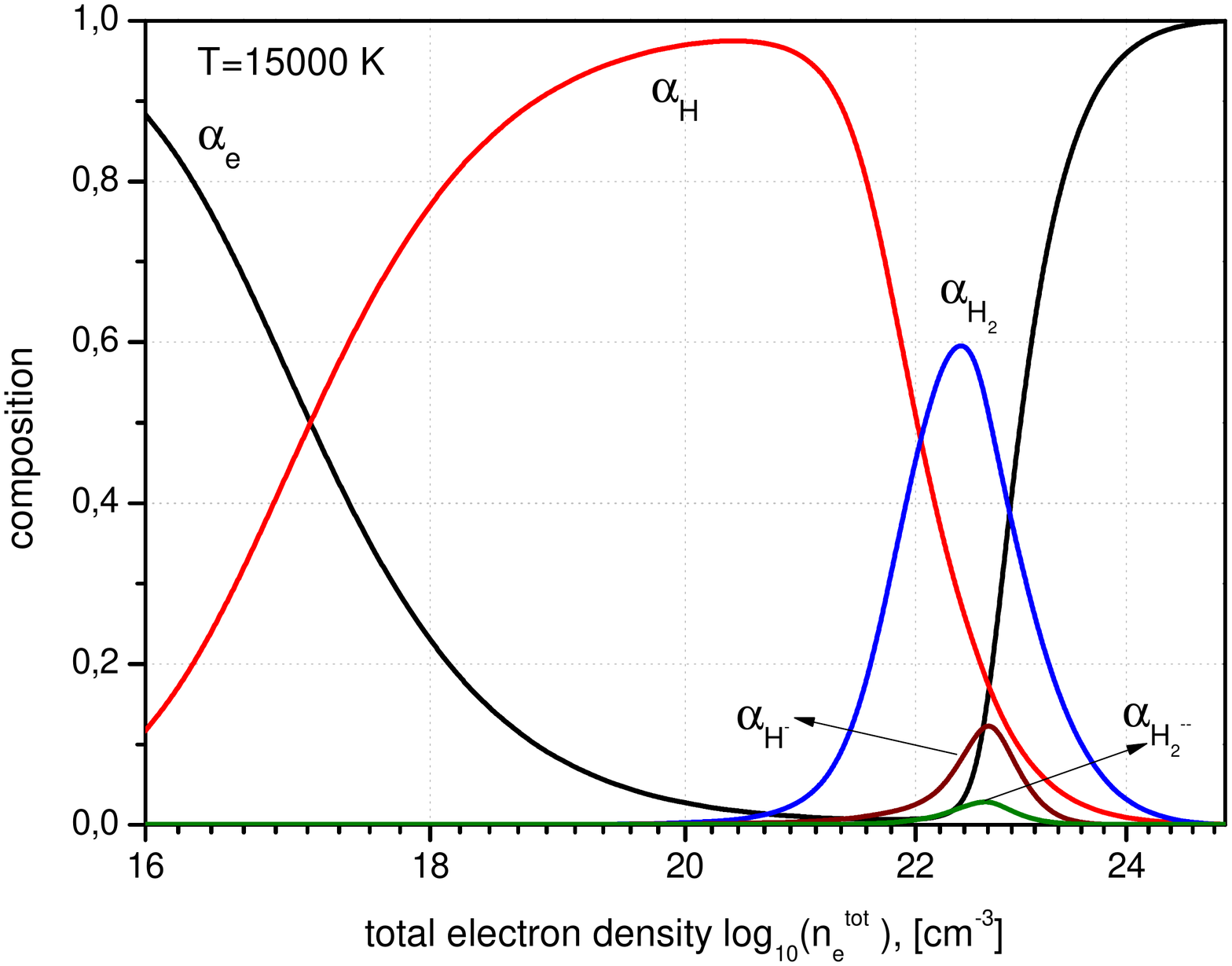}
\caption{(Color online) The composition of hydrogen plasmas for $T=15000$ K}
\label{fig:composition}  
\end{figure*}

We study partially ionized hydrogen plasma with electrons ($e$), ions (protons $i$), hydrogen atoms ($\rm H$), and
hydrogen molecules ($\rm H_2$) as constituents starting from the chemical picture. We focus on the interaction of $e-\rm H_2$ and 
its contribution to the equation of state. The influence of the molecular component
is essential in dense partially ionized plasmas. 
The composition of hydrogen plasma has been calculated following a set of mass action laws \cite{KSK}. 
Fig.\ref{fig:composition} presents a standard approach \cite{KSK} to the composition of hydrogen plasma, exemplarily for $T=15000$ K, 
with fractions $\alpha_c=Z_c n_c/n_e^{\rm tot}$, $n_c$ is density of species, 
$n_e^{\rm tot}$ is total electron density, $Z_c$ is the number of electrons in the corresponding bound states.      
The free electron fraction is decreasing until total electron densities of
$10^{23}$ cm$^{-3}$ before pressure ionization sets in.    
The fraction of hydrogen atoms is dominating in the density region $\approx$ $10^{18}-10^{24}$ cm$^{-3}$. At densities above 
$10^{21} ~\rm cm^{-3}$, the molecule fraction plays an essential role in physical processes.  Note, that following \cite{KSK} the interactions 
between electrons and clusters are taken into account 
by a hard-core model to calculate the composition 
in Fig.\ref{fig:composition}. The use of experimental data for the interaction parts of chemical potentials can give more accurate data for the composition. 
In the present work, we consider partially ionized 
hydrogen plasmas at temperatures $T\leq10^{5}$ K 
and densities up to $10^{22}$ cm$^{-3}$ until degeneracy effects play an essential role. As a new ingredient, the contribution of scattering states is considered.
We apply the cluster virial expansion approach to study the contribution of the electron-molecule interaction to thermodynamical properties.      
The present work is organized as follows: In Section \ref{sec:CVE}, we
briefly review the cluster-virial expansion and the Beth-Uhlenbeck formula for the second virial coefficient. Section
\ref{sec:SCD} contains the calculation of the scattering phase shifts for
the electron-molecule system, both via experimental differential
cross sections and phase shifts from appropriate pseudopotentials. In Section \ref{sec:SVC}, the phase shifts are used to calculate
the corresponding second virial coefficient. Results for the ${\rm H_2}-e$ second virial coefficient are given for different temperatures, 
and consequences for the composition are considered. 
Conclusions are drawn in Section \ref{sec:Con}. 


\section{Cluster virial expansion and Beth-Uhlenbeck formula\label{sec:CVE}}
The cluster virial expansion for the equation of state \cite{Huang} can be written as 
function of fugacities, $z_c=e^{\beta(\mu_c-E_c^{(0)})}$,
\begin{equation}
 \beta p = \sum\limits_c\frac{2s_c+1}{\Lambda_c^3}\big(z_c+\sum\limits_d z_c z_d \tilde b_{cd}+\cdots\big), 
\label{eq:druck}
\end{equation}
where $c$ denotes species ($c=e,i, \rm H, \rm H_2$), $s_c$ - spin,  $\mu_c$ - chemical potential, $E_{cd}^{(0)}$ - binding energy for isolated cluster species,
$\Lambda_c=(2\pi\hbar^2/k_BTm_c)^{1/2}$ - the thermal wavelength of species $c$. 
The first term is the ideal part of the pressure. 
The contribution of the Coloumb interaction between charge particles and the interaction with neutrals must be treated differently. 
For the Coloumb interactions ($e-e$, $e-i$, $i-i$), the non-ideal contributions of the equation of state have been intensively 
investigated, for a review see Ref. \cite {KKER}. For the interaction with neutrals the dimensionless second virial coefficient $\tilde b_{cd}$ 
is determined by the respective interactions of 
$e-\rm{H}$, $i-\rm{H}$, $\rm{H}-\rm{H}$, $e-\rm{H_2}$, $i-\rm{H_2}$, $\rm{H_2}-\rm{H_2}$, $\rm{H}-\rm{H_2}$  pairs.         
In particular, $\tilde b_{\rm HH}$ was calculated in Refs. \cite{RN} and $\tilde b_{e\rm H}$ was studied in Ref. \cite{OFRR}. 

An exact quantum mechanical expression for 
the second virial coefficient was given by Beth and Uhlenbeck \cite{Huang}:
\begin{eqnarray}
\tilde b_{cd}=\sum_{\ell}^{}(2\ell+1)
\sum_{n} (\mathrm{e}^{-\beta E_{cd}^{n\ell}}-1)
\nonumber\\
+\sum_{\ell=0}^{\infty}(2\ell+1)
 \frac{\beta}{\pi}\int_{0}^{\infty}\mathrm{e}^{-\beta E}\,\eta_\ell^{cd}(E)\,dE,
\label{eq:BU}
\end{eqnarray}
 where $\ell$ is orbital momentum, $\eta_\ell^{cd}(E)$ is the scattering phase shift, $E$ is the energy of incident particles, $E_{c}^{n\ell}$ is binding energy of the 
 state with quantum numbers $n\ell$. 
The first term is the bound part $\tilde b_{cd}^{\rm bound}$ and the second is the scattering part $\tilde b_{cd}^{\rm sc}$.   
  
In this paper we focus on the second virial coefficient for $e-\rm H_2$. 
We calculate the scattering part of the second virial coefficient $\tilde b_{{\rm H_2}e}$ due to electron and hydrogen molecule interaction.
The bound part includes a new component  $\rm H_2^{-}$ in the system. The calculation of the bound part requires 
the binding energy of the negative hydrogen molecule, which was taken from the literature \cite{harcourt}. Note that alternatively to Eq.(\ref{eq:BU}),  
the bound state $\rm H_2^{-}$ can be considered as a new species $c$ in calculating thermodynamic properties. 


\section{Scattering data\label{sec:SCD}}

\begin{figure*}[htp]
\includegraphics[width=0.65\textwidth]{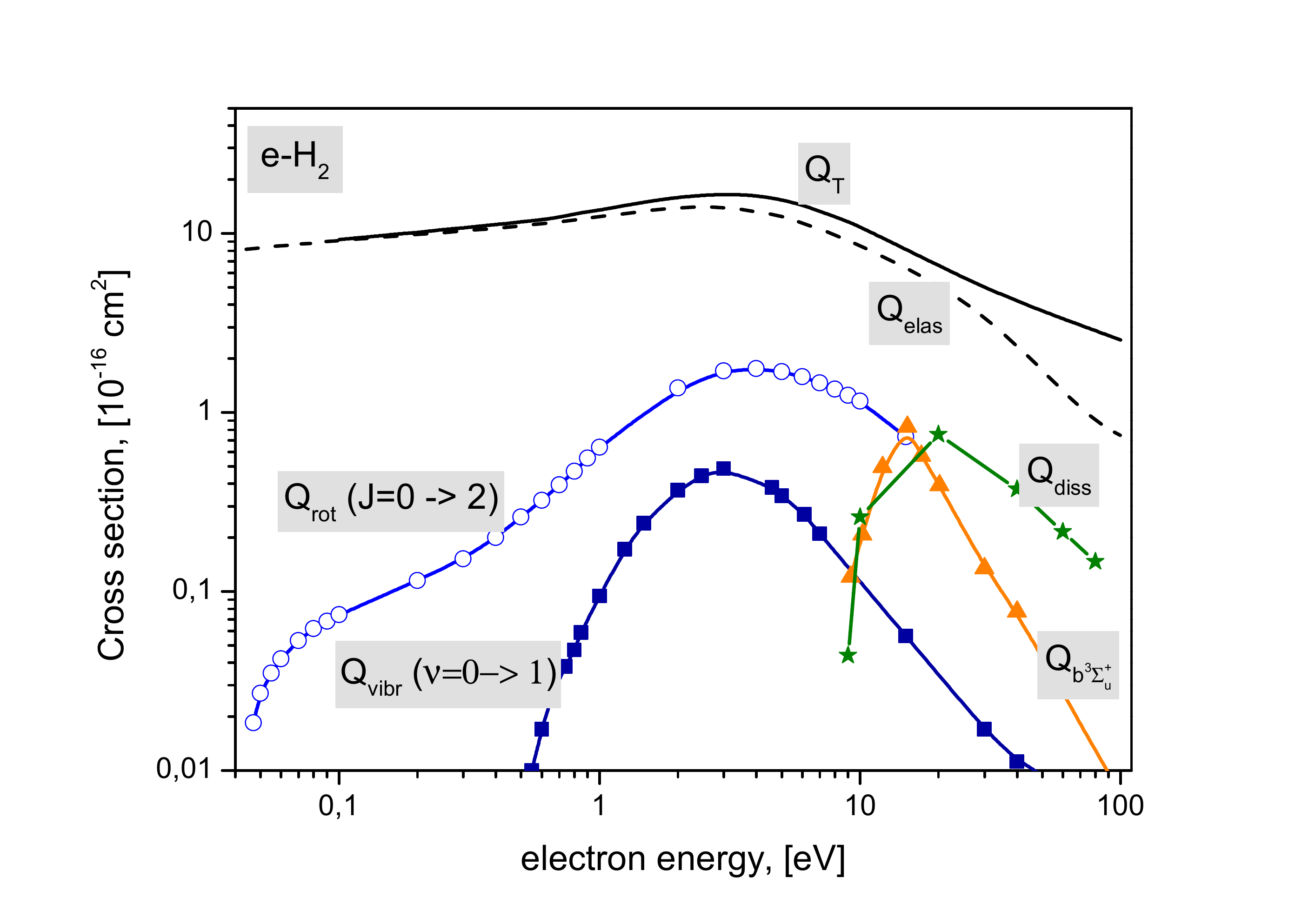}
\caption{(Color online) Cross section for $e$-$\rm H_2$ collision. Adapted from Ref. \cite{itikawa}}
\label{fig:1} 
\end{figure*}

Scattering phase shifts data for $e-\rm H_2$ can be employed to calculate the
second virial coefficient $\tilde b_{{\rm H_2} e}$ using the Beth-Uhlenbeck formula (\ref{eq:BU}).
Due to internal degrees of freedom of the $\rm H_2$ molecule,
the $e - \rm H_2$ scattering problem is more complex compared to the $e - \rm H$ system.
 In addition to electronic excitation and ionization, other processes at energies below the ionization limit $E_i=124417.49~{\rm cm^{-1}} \approx 15.42~{\rm eV}$ \cite{LSH} 
 have to be considered.  
It is clear that the total cross section $Q_{\rm T}$ includes all these processes:
\begin{eqnarray}
Q_{\rm T}=Q_{\rm elas}+Q_{\rm att}+Q_{\rm diss} +\sum Q_{\rm excit}, 
\label{eq:Q1} 
\end{eqnarray}
where cross sections for elastic scattering is $Q_{\rm elas}$, for dissociative attachment is $Q_{\rm att}$, for impact dissociation is $Q_{\rm diss}$.     
$\sum Q_{\rm excit}$ is the sum of all excitation cross sections of rotational, vibrational, and electronic states. 
Table \ref{tab:1} shows the contribution of those transitions, which were estimated in Ref.\cite{TTR}.
\begin{table*}[htp]
\centering
\caption{\label{tab:1} Contribution of various transitions to total $e-{\rm H_2}$ scattering cross sections [$a_{\rm B}^2$] at different impact energies}
 \begin{ruledtabular}
 \begin{tabular}{l c c c c c c c}
Transition $\setminus$ energy [eV] & ~7 &  ~10 & ~13.6 & ~20 & ~45 & ~60 & ~81.6\\
\hline
A. Total scattering \cite{GBS} &~ 42.1&~ 33.7&~ 26.9&~ 20.0&&& \\
B. Electronic excitation \cite{TTR}&             ~0.0 & ~0.73 &~ 3.27 & ~5.23&&&\\
C. A minus B                        & ~42.1& ~33.0& ~23.6& ~14.8&&&\\
D. Elastic scattering \cite{TTR}               &~41.5&~32.6&~23.4& ~14.7&~ 7.86&~5.97&~4.44\\
E. Vibrational excitation \cite{TTR}           &~0.60& ~0.35&~ 0.11&~ 0.05&~0.05&~0.03&~0.03\\
\end{tabular}
\end{ruledtabular}
\end{table*}
The rotational excitation is not given. However, this channel is mixed into elastic and vibrational 
excitation transitions. As can be seen in Table \ref{tab:1}, the electronic excitation channel is closed until 7 eV.   
The vibrational transition is about $1.4\%$ at 7 eV of total cross section and it is decreasing with increasing incident energy. 
Up to 10 eV the total cross section is dominated by elastic contributions. This statement is further verified by considering the collision cross sections, see Figure \ref{fig:1}. 
The total cross section $Q_{\rm T}$ is obtained from beam measurements and was determined as recommended value in Ref. \cite{itikawa}.
 The rotational excitation channel is already open 
at $44 \cdot 10^{-3}$ eV for the lowest rotational state ($J=0\rightarrow 2$). The vibrational channel sets in at  $0.516$ eV, the 
electronic excitation channel at $7$ eV. The contributions
of electronic, rotational, and vibrational excitations to the total cross
section below $10$ eV are not more than 1.04\%, 10\%, and 2.89\%, respectively. Note
that $Q_{\rm rot}$ is determined from theoretical calculations \cite{Morrison}. According to
these estimations, the $e - \rm H_2$ scattering process below $10$ eV is determined
with sufficient accuracy by the elastic contribution only.

\subsection{$e-\rm H_2$ scattering theory\label{sec:SF}}

Various theoretical methods were developed to solve the Schr\"odinger equation for $e-\rm H_2$ scattering process. The T-matrix expansion method, 
the Schwinger variational method, the R-matrix method  are so called basis-set expansion methods applied for electron-molecule system (see the detailed review of theoretical 
methods in Ref. \cite{Lane}). An alternative approach is an one-particle picture for the 
description of the elastic scattering process in fixed-nuclei (Born-Oppenheimer) limit \cite{Lane}.  
Molecules are fixed in position and the Schr\"odinger equation is solved for the electrons 
in the static electric potential arising due to the molecular configuration.

In this paper, we use the simplest approximation to solve the scattering problem of electron-hydrogen molecule system. 
We assume that molecules are fixed in space and it is not rotating and not vibrating. 
In this case, the interaction between an electron and a molecule is treated similar to that of an electron-atom system. That means the electron is 
scattered by a optical potential  $V$
\begin{eqnarray}
{\rm H}_{\rm eff}=T_e+V,  
\label{eq:S3} 
\end{eqnarray} 
and we use the phase function method \cite{calogero, babikov} to solve the Schr\"odinger equation.

The phase function equation or so called Calogero equation for the scattering phase shift $\eta_{\ell}$ is 
\begin{eqnarray}
\frac{d\eta_{\ell}^{cd}(k,r)} {dr}=&-\frac{1}{k}U(r)[\cos\eta_{\ell}^{cd}(k,r)J_{\ell}(k,r)
\nonumber\\
&-\sin\eta_{\ell}^{cd}(k,r)n_{\ell}(k,r)]^2
\label {eq:C}
\end{eqnarray}
(see more details in Ref. \cite{yultuz_cpp}).
The Calogero equation has an initial condition $\eta_{\ell}^{cd}(k,0)=0$, where $k$
is the wave number, $\ell$ are orbital quantum
numbers, $J_{\ell}(k,r)$and $n_{\ell}(k,r)$ are the Riccati-Bessel functions, $U(r)=\frac{2m_{cd}}{\hbar^2}V(r)$, $V(r)$ is the interaction potential.
The energy-dependent scattering phase shifts $\eta_{\ell}^{cd}(k)$ are determined as
$\eta_{\ell}^{cd}(k)=\lim_{r\to\infty}\eta_{\ell}^{cd}(k,r)$.

\subsection{Interaction potential}
As we mentioned above, the accurate calculation 
of the scattering problem requires an adequate approximation of the optical potential.  It is a full projectile-target (electron-molecule)
 interaction potential which consists of static, exchange and polarization contributions.
The static potential is given by the electrostatic interaction between the projectile and the constituent particles of the target \cite{Morrison}. 
The exchange effect is important at low energies, it occurs due to indistinguishability of the projectile and target electrons. 
The polarization potential describes induced distortions of the target by the impact electron.  
Since the goal is to consider a collision process in plasma, the last effect (polarization) is particularly important for the description of plasma properties. 
 Collisions of electrons on molecules in plasmas with not too high densities occurs at large distances.  
The polarization potential for the electron-atom interaction has the asymptotic behavior $\alpha/2r^4$ (at large distances) with the polarizability $\alpha$ of the atom. 
Since this potential is diverging at small distances, the Buckingham potential was suggested for $e-a$ interaction \cite{RRZ}:   
\begin{eqnarray}
V_{ea}=-\frac{\alpha}{2(r^2+r_0^2)^2}, 
\label{eq:V3} 
\end{eqnarray}     
where $r_0$ is a cut-off radius. For hydrogen atoms $\alpha=4.5~a_{\rm B}^3$ and $r_0=1.456~a_{\rm B}$ \cite{RedPR}. If we consider interaction of electrons with diatomic molecules, 
the polarization model is modified \cite{Lane,Morrison}:
\begin{eqnarray}
V_{e\rm H_2}(r)=\Big(-\frac{\alpha_0}{2r^4}-\frac{\alpha_2}{2r^4}P_2(\cos\theta_p)\Big)
\nonumber\\
\times\Big(1-\exp{(r/r_0)^6}\Big),
\label{eq:V2} 
\end{eqnarray}  
where $\alpha_0$ and $\alpha_2$ are polarizabilities
parallel and perpendicular to the internuclear axis $\vec e_R$, respectively. $P_2(\cos\theta_p)$ is the Legendre polynomial. 
$\theta_p$ is the angle between the direction of the incident electron and the $z$ - axis.
This potential describes the interaction of molecule positioned at the origin and the $z$ -
 axis coincides with $\vec e_R$.    
For $\rm H_2$ (internuclear distance $R=1.4~a_{\rm B}$), we use the experimental data of polarizabilities 
$\alpha_0=5.4265~a_{\rm B}^{3},~ \alpha_2 =1.3567~a_{\rm B}^{3}$ \cite{Morrison}.

\subsection{Phase shifts}
\begin{figure}
\includegraphics[width=0.45\textwidth]{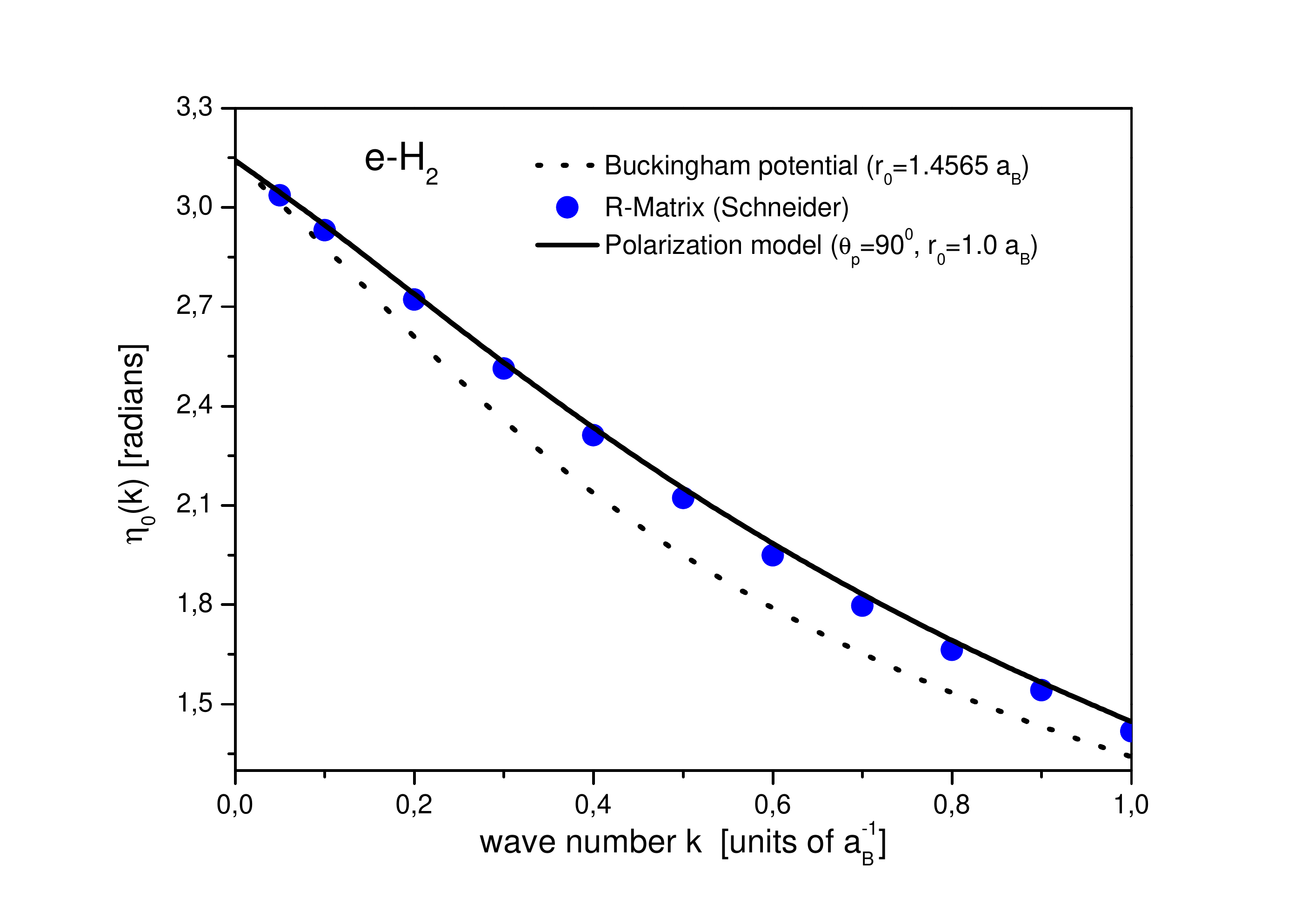}
\caption{(Color online) s-wave scattering phase shifts for $e-{\rm H_2}$. R - matrix data \cite{schneider} 
is compared with the results of the present calculation on the basis of the polarization models (\ref{eq:V2}) and (\ref{eq:V3})}.
\label{fig:swave} 
\end{figure}

\begin{figure}
\includegraphics[width=0.45\textwidth]{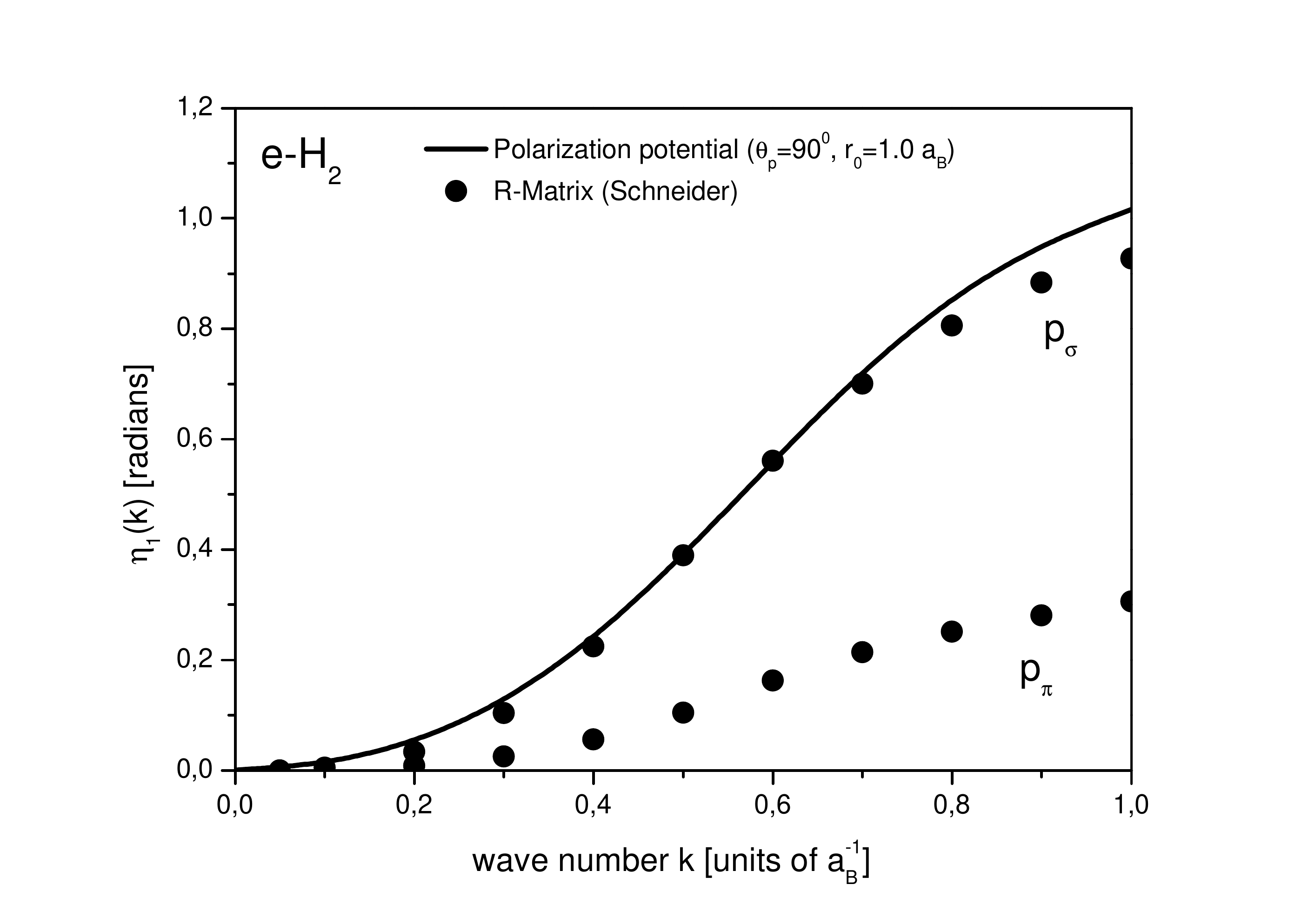}
\caption{p-wave scattering phase shifts for $e-{\rm H_2}$. R - matrix data \cite{schneider} 
is compared with the results of the present calculation on the basis of the polarization model (\ref{eq:V2})}
\label{fig:pwave} 
\end{figure}

\begin{figure}
\includegraphics[width=0.45\textwidth]{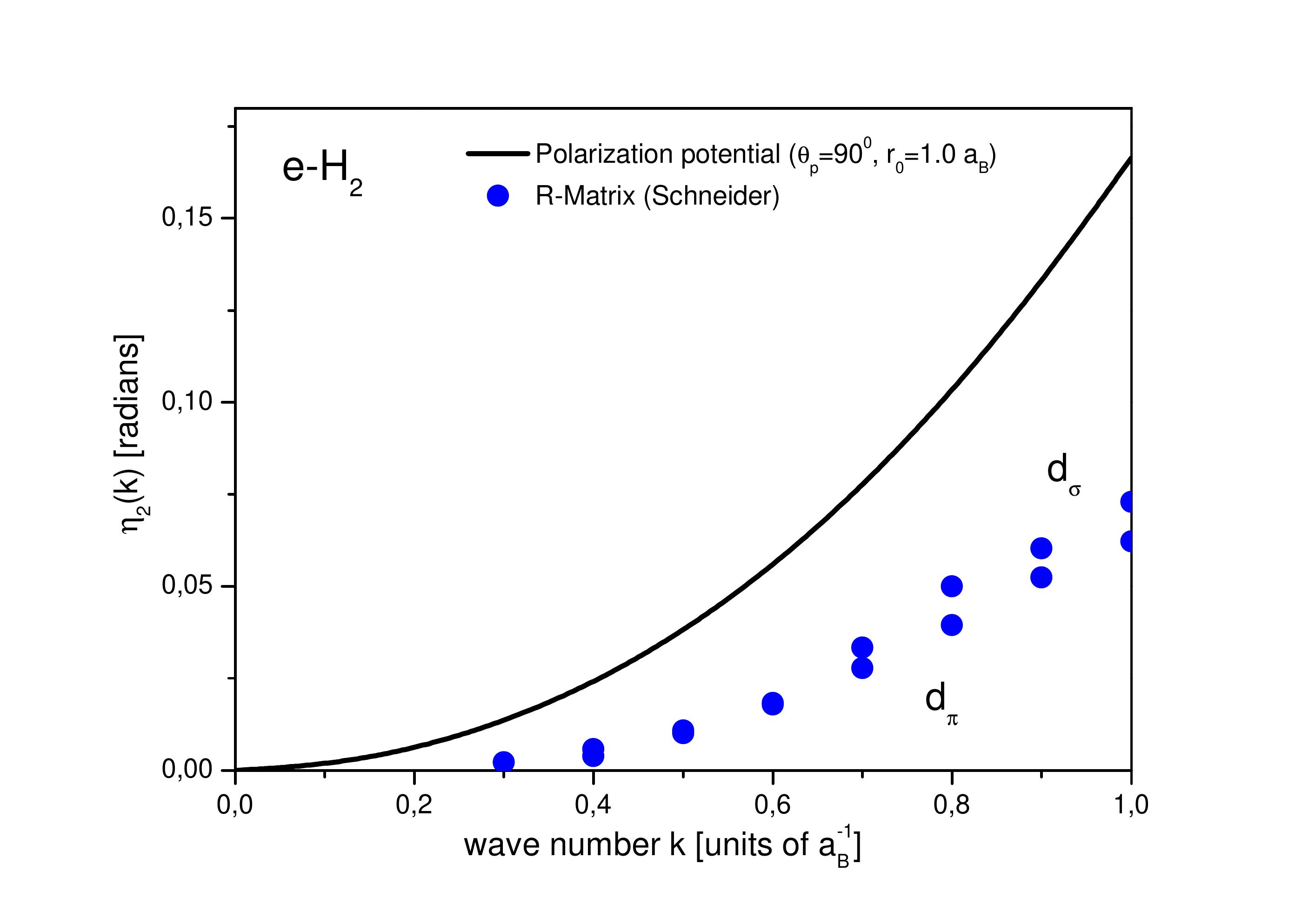}
\caption{(Color online) d-wave scattering phase shifts for $e-{\rm H_2}$. R - matrix data \cite{schneider} 
is compared with the results of the present calculation on the basis of the polarization model (\ref{eq:V2})}
\label{fig:dwave} 
\end{figure}

The solution of the Calogero equation is used to obtain the scattering phase shifts. 
The results for different orbital momenta on the basis 
of the polarization model (\ref{eq:V2}) are shown in the Figures \ref{fig:swave}, \ref{fig:pwave} and \ref{fig:dwave}. 
We consider the phase shifts results in Fig.\ref{fig:swave}. At $k=0$ the s-wave scattering phase shifts $\eta_0(0)$ tends to the value of $\pi$. According to the 
Levinson theorem \cite{levinson} $\eta(0)=n\pi$ (where $n$ is the number of bound states), it corresponds to one bound state. In our case it is the negative hydrogen molecule.   
$\rm H_2^{-}$ is a metastable state, that appears in reactions like dissociative attachment (${\rm H_2} +e\rightarrow {\rm H_2^{-}}\rightarrow {\rm H}+{\rm H^{-}}
$) and associative attachment (${\rm H}+{\rm H^{-}}\rightarrow {\rm H_2^{-}}\rightarrow 
{\rm H_2} +e$). The stability of this state is discussed in the literature.  Recently, the lifetime of this metastable state was measured as $~ 5-8~\mu \rm s$ \cite{SSNHMI}. 
The theoretical value of electron affinity (or binding energy) for the bound state $\rm H_2^-$ is $2.08\cdot10^{-2}$ eV/atom corresponding to $2 ~\rm kJ/mole$ \cite{harcourt}. 
This value has been taken to calculate the bound part of the second virial coefficient in the Beth-Uhlenbeck formula (\ref{eq:BU}).

The phase shifts data for the ${\rm s}$-channel are compared with the R-matrix data of Ref. \cite{schneider}. 
As one can see, the present results for the polarization model (\ref{eq:V2}) have good agreement with Schneider's data. 
The parameter values of 
the polarization potential (\ref{eq:V2}) $r_0$ and $\theta_p$ are fit to get a good agreement in phase shifts with theoretical data.  
The cut-off radius in this calculation is taken as $r_0=1.0~a_{\rm B}$, and $\theta_p=90^\circ$. 
Note, that no direct measurements of the phase shifts 
can be found in literature, only scattering cross section data. 

In Figures \ref{fig:pwave} and \ref{fig:dwave} the scattering phase shifts for $\ell=1, ~2$ are presented using the same parameters for $r_0$ and $\theta_p$ as for s-wave. 
Both phase shifts are zero at zero incident energy of the electron, since no bound states exist for these scattering channels. 
The d-wave results are very small in comparison with the s channel at low energy limit. In general, to calculate the second virial 
coefficients the phase shifts for $\ell < 3$ are enough to obtain accurate results. 
The comparison of ${\rm p}$- and ${\rm d}$ - waves with Schneider's data \cite{schneider} shows deviations. This can be explained 
by the different methods we used and neglection of symmetry effects in our approach. 
In the present calculation, a molecule without structure is considered, so the comparison of data with other theoretical works in $\sigma$ and $\pi$ 
orbitals is not possible. The only confirmation of this studies can be the comparison of calculated differential cross section with experimental data.

\subsection{Elastic differential cross section}
Experimental data for the electron-molecule collisions were collected by 
Trajmar \cite{TRC}, Brunger \cite{BB} and Itikawa \cite{itikawa}. As was discussed above, 
rotational excitation channels are already open at a very low energy; for instance, the lowest rotational state energy($J=0\rightarrow 2$) is   $44.13 \cdot 10^{-3}$ eV. 
Therefore, the experimental data for elastic differential cross section include rotational excitations and are rotationally summed. 
Only in the experiment by Linder and Schmidt \cite{linder}, elastic scattering was separated from rotational excitation. So, the data of Linder and Schmidt were 
taken to compare this studies' results for differential cross sections. Also recent experimental data from Muse et al. \cite{muse} are taken 
to perform the comparison of differential cross sections.

\begin{figure}
\includegraphics[width=0.45\textwidth]{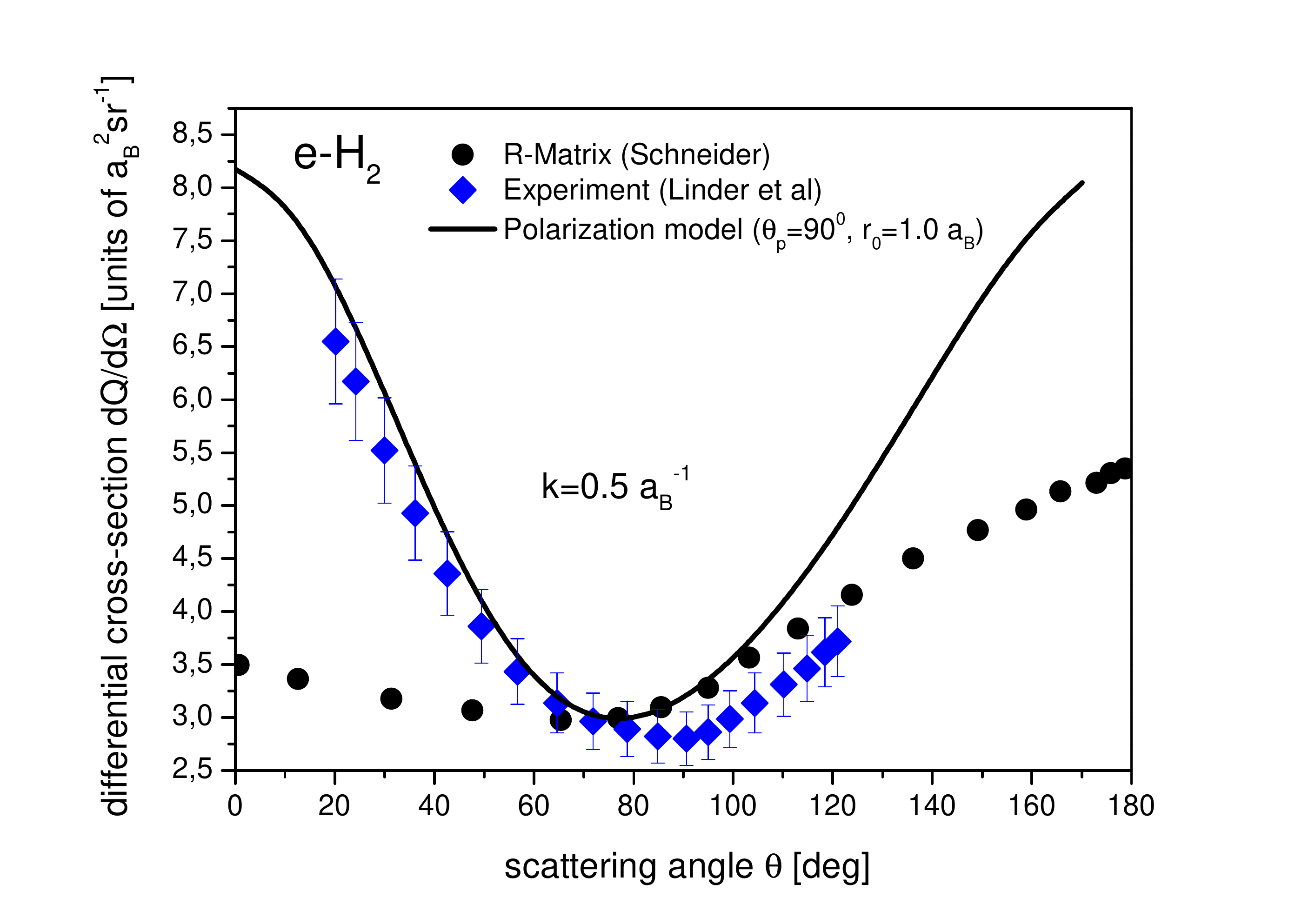}
\caption{(Color online) Differential cross sections for $e-{\rm H_2}$ at wave number $k=0.5 a_{\rm B}^{-1}$. 
Solid line shows calculation with the polarization model (\ref{eq:V2}); square line is experimental data Ref.\cite{linder}; 
triangle line is theoretical data Ref.\cite{schneider}}.
\label{fig:DCS_k} 
\end{figure}

\begin{figure}
\includegraphics[width=0.45\textwidth]{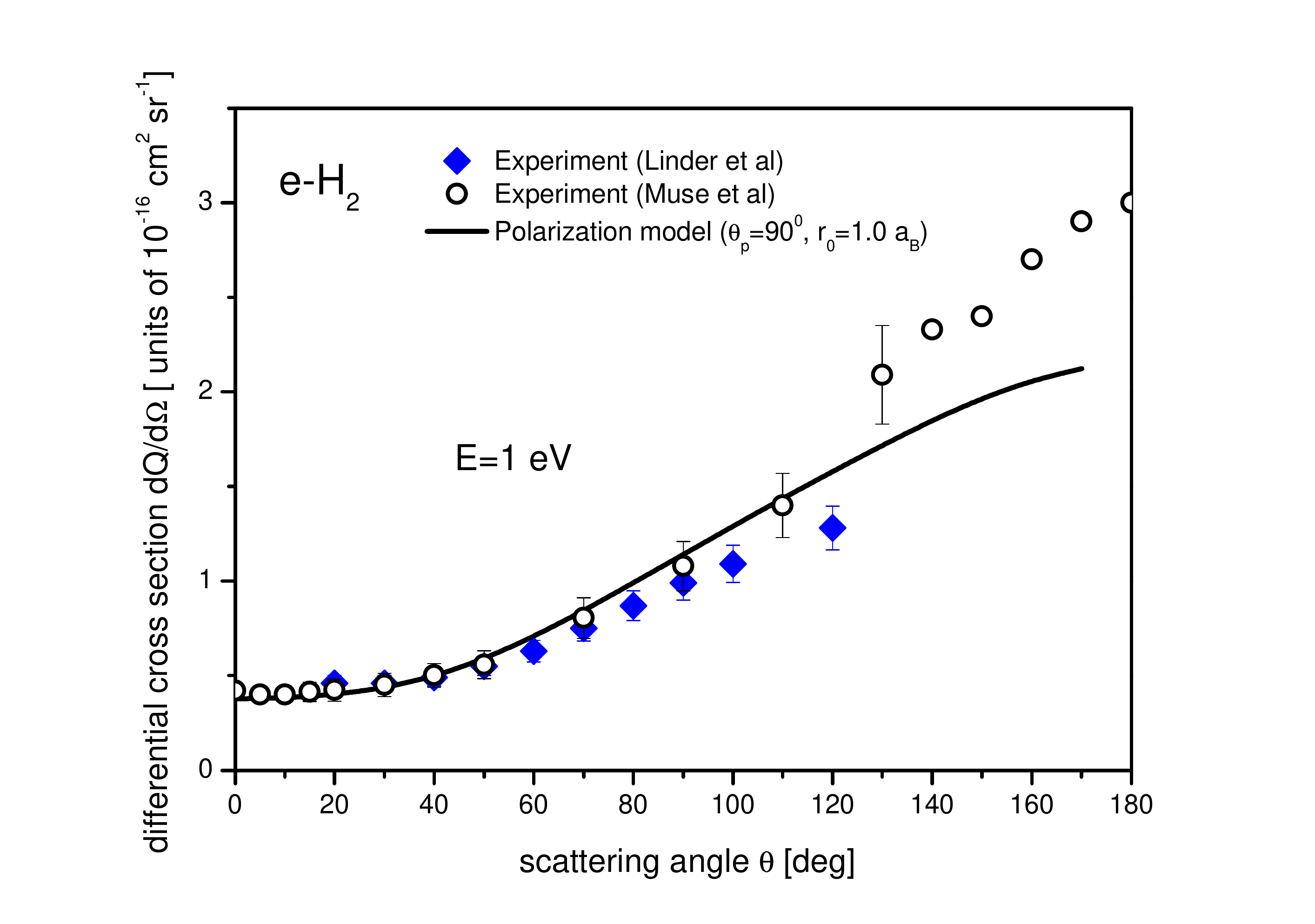}
\caption{(Color online) Differential cross sections for $e-{\rm H_2}$ at incident energies 1 eV. Solid line presents calculation with the polarization model (\ref{eq:V2});
 square line is experimental data Ref.\cite{linder}; circle - experimental data Ref. \cite{muse}}
\label{fig:DCS1} 
\end{figure}

\begin{figure}
\includegraphics[width=0.45\textwidth]{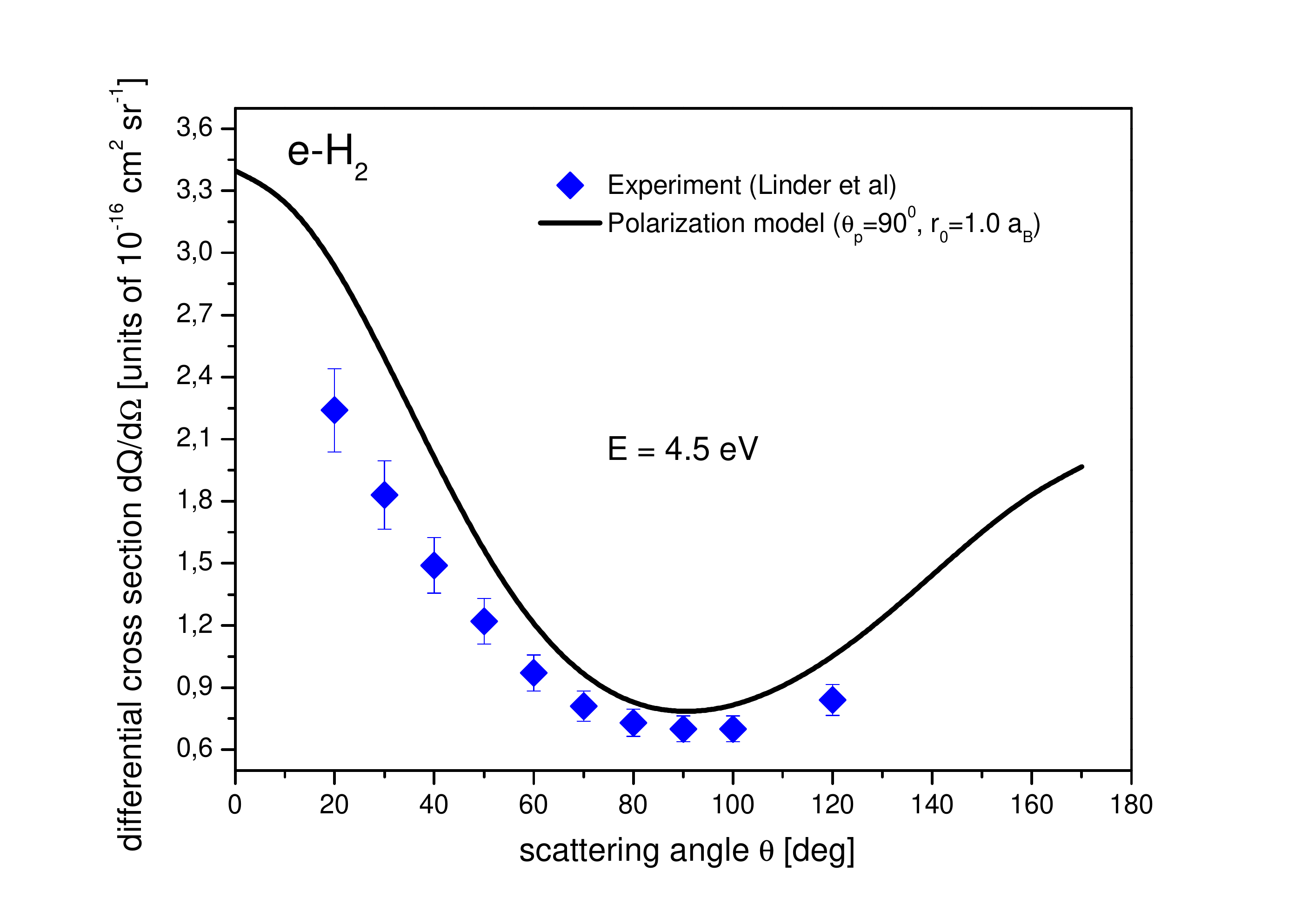}
\caption{(Color online) Differential cross sections for $e-{\rm H_2}$ at incident energies 4.5 eV. Solid line presents calculation with the polarization model (\ref{eq:V2});
 square line is experimental data Ref.\cite{linder}}
\label{fig:DCS45} 
\end{figure}

\begin{figure}
\includegraphics[width=0.45\textwidth]{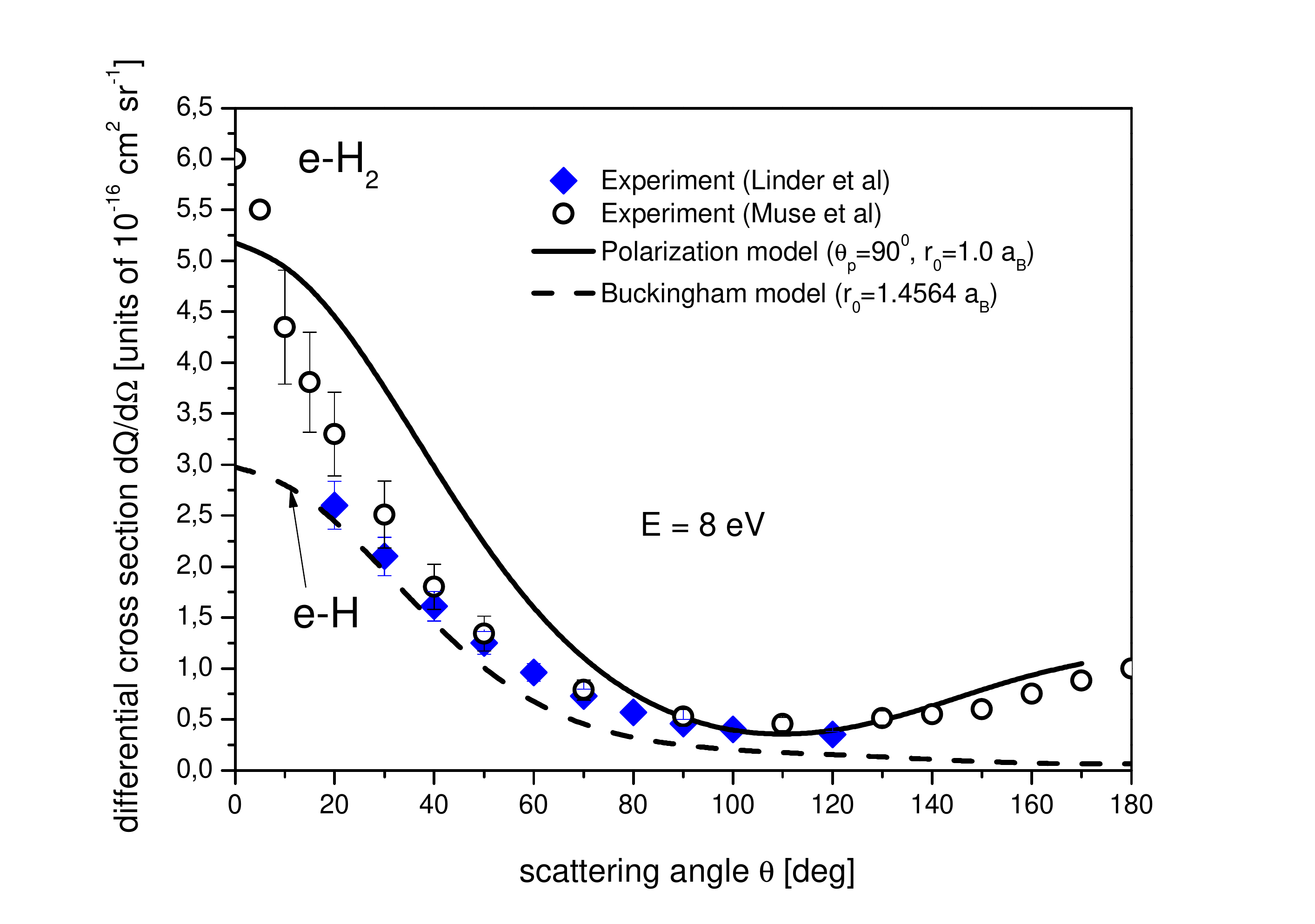}
\caption{(Color online) Differential cross sections for $e-{\rm H_2}$ at incident energies 8 eV. Solid line presents calculation with polarization model (\ref{eq:V2});
 square line is experimental data Ref.\cite{linder}; circle - experimental data Ref. \cite{muse}; dashed line - present calculation with the polarization model 
 (\ref{eq:V3})}
\label{fig:DCS8} 
\end{figure}

Using the obtained phase shifts, the differential cross sections can be calculated by the following formula:
\begin{eqnarray}
\frac{dQ(k,\theta)}{d\Omega}= \left\vert\frac{1}{2ik}\sum\limits_{\ell}(2\ell+1)
[e^{2i\eta_{\ell}^{cd}(k)}-1]P_{\ell}(\cos\theta)\right\vert^2,
\nonumber\\
\label{eq:DCS}
\end{eqnarray}
where $\theta$ is the scattering angle (do not confuse with $\theta_p$). In our calculation we include the orbital momentum until $\ell =5$. 
The dependence of the differential cross section on scattering angle is shown in Figs. \ref{fig:DCS_k}, \ref{fig:DCS1}, \ref{fig:DCS45} and \ref{fig:DCS8} for different 
incident energies of the electron. In Fig. \ref{fig:DCS_k} the results of the polarization model (\ref{eq:V2}) is compared with R-matrix data of Schneider \cite{schneider} 
and experimental data \cite{linder} at $k=0.5~a_{\rm B}^{-1}$. Our results describe better the collision process at small scattering angles, whereas the R-matrix data 
works only at middle angles.        
The comparison of our results with experimental data \cite{linder,muse} at other energies shows a good agreement almost at all scattering angles. With increasing 
incident energy, slight deviation between experiment and our calculation is observed. It can be explained by an increasing contribution of rotational excitations, 
which is not included in our calculation.  In Fig.\ref{fig:DCS8} also the differential cross section for electron and atom scattering is presented. 
The atomic cross section, calculated using the Buckingham potential (\ref{eq:V3}), is smaller than the molecular cross section almost by a factor of 2.  
Although we use a simple 
approximation to describe the scattering process between electron and hydrogen molecule, our results are reliable which was shown by comparison with experimental data. 


\section{Results and discussion\label{sec:SVC}}
\subsection{Second virial coefficient for $e-\rm H_2$ interaction}

The data of scattering phase shifts shown in the \Figref{swave, pwave, dwave} which 
are based on experimental data, will be used for calculations of the second virial coefficient using the
Beth-Uhlenbeck formula (\ref{eq:BU}). The phase shifts are obtained using the
polarization model Eq. (\ref{eq:V2}). Table \ref{tab:3} shows results for the normalized second
virial coefficients $\tilde b_{{\rm H_2} e}^{\rm sc}$ and $\tilde b_{{\rm H_2} e}^{\rm bound}$ for the scattering and the bound parts, respectively. 
The second, third and fourth columns of the table present
data for the contribution of s, p, and d-waves to the scattering
part of the second virial coefficient, respectively. Higher order contributions are small
and negligible for the temperature range considered here. The $\tilde{b}_{{\rm H_2} e}^{\rm sc}$ for higher orbital momenta 
is weaker than for the $\rm s$-wave. 
With increasing temperature, the scattering part of the second virial coefficient for the $\rm s$-wave 
decreases weakly, but the results for the other two channels increase. 
This occurs due to the difference in behaviour of  phase shifts, see \Figref{swave, pwave, dwave}. 
The sixth column shows the bound part ($\rm H_2^-$) of the second virial coefficient.  
The bound part is strongly depending on temperature. At low temperatures bound particles (clusters) 
are important.  The full second virial coefficient is presented in the last column. 
The dependence of the full second virial 
coefficient on temperature is determined by the scattering part. 
Data from Table \ref{tab:3} can be used to study thermodynamical properties 
of the system. Note, that the second virial coefficients do not depend on the density of the plasma. The density dependence is included in the virial expansions for 
thermodynamical functions, see for instance the pressure, Eq.(\ref{eq:druck}).  
\begin{table*}[htp]
\center
\caption{\label{tab:3}Scattering and bound part of the second cluster virial coefficient $\tilde b_{{\rm H_2} e}$ for different temperatures.}
\begin{ruledtabular}
\item[]
\begin{tabular}{ccccccc}
$T$, [K]&$\tilde b_{{\rm H_2} e}^{\rm sc}$, s-wave&$\tilde b_{{\rm H_2} e}^{\rm sc}$, p-wave&$\tilde b_{{\rm H_2} e}^{\rm sc}$, d-wave
&$\tilde b_{{\rm H_2} e}^{\rm sc}$, full&$\tilde b_{{\rm H_2} e}^{\rm bound}$&$\tilde b_{{\rm H_2} e}$, full\\
\hline 
5000& 0.9308& 0.0179&0.0034&0.9522&0.0494&1.0015\\
7000& 0.9173& 0.0260&0.0049&0.9499&0.0350&0.9832\\
8000& 0.9113& 0.0301&0.0056&0.9482&0.0306&0.9776\\
9000& 0.9056& 0.0343&0.0064&0.9464&0.0271&0.9734\\
10000&0.9002& 0.0386&0.0071&0.9460&0.0244&0.9703\\
11000&0.8951& 0.0429&0.0079&0.9460&0.0222&0.9682\\
12000&0.8903& 0.0473&0.0086&0.9462&0.0203&0.9665\\
13000&0.8856& 0.0517&0.0094&0.9468&0.0187&0.9656\\
14000&0.8813& 0.0562&0.0101&0.9476&0.0173&0.9649\\
15000&0.8770& 0.0607&0.0108&0.9485&0.0162&0.9647\\
20000&0.8576& 0.0836&0.0146&0.9560&0.0121&0.9681\\
\end{tabular}
\end{ruledtabular}
\end{table*}

\subsection{Ionization Equilibrium}
The interaction between electrons and neutral clusters can play an essential role for the plasma composition. 
In a system with charged particles, hydrogen atoms and hydrogen molecules, the following chemical reactions are possible: 
$ e+i\leftrightharpoons\rm H $, $\rm H+\rm H \leftrightharpoons \rm H_2$, $e+\rm H\leftrightharpoons \rm H^-$ and $e+\rm H_2\leftrightharpoons \rm H^-_2$.
Each reaction corresponds to a chemical equilibrium with respect to the chemical potentials, respectively:
\begin{equation}
\begin{aligned}
\mu_e+\mu_i &= \mu_{\rm H}, 
\\
\mu_{\rm H}+\mu_{\rm H} &= \mu_{\rm H_2},  
\\
\mu_e+\mu_{\rm H}&= \mu_{\rm H^-},
\\
\mu_e+\mu_{\rm H_2}&= \mu_{\rm H^-_2}. 
\end{aligned}
\label{eq:s1}
\end{equation}
Note that the clusters $\rm H^-$ and $\rm H_2^-$
are included now as components in the chemical picture. Alternatively, the contribution of these bound states can also be obtained consistently 
from the Beth-Uhlenbeck formula (\ref{eq:BU}) as bound state contribution. Further possible cluster states are not considered in this work, in particular the $\rm H_2^+$ bound state 
will be obtained in the ${\rm H}-i$ interaction channel.

In non-ideal plasma the chemical potential can be divided into ideal and non-ideal parts. For instance, for the chemical potential, 
the following expression defines the virial coefficients \cite{Huang}
\begin{equation}
\mu_c=\mu_c^{\rm id}-k_{\rm B}T\Big(2\sum_d n_d b_{cd}+3\sum_{de}n_dn_eb_{cde}+...\Big), 
\label{eq:mu}
\end{equation}
where $b_{cd}$ is the second virial coefficient. 
There is a connection between the dimensionless second virial coefficient $\tilde b_{cd}$ (see Eq.(\ref{eq:druck})) and  $b_{cd}$: 
$\tilde b_{cd}=b_{cd} g_d/\Lambda_d^3$, where $g_d$ is the spin degeneracy factor.    
The first term of Eq.(\ref{eq:mu}) $\mu_c^{\rm id}=k_{\rm B}T\ln(\frac{n_c\Lambda_c^3}{g_c})+E_c^{(0)}$ is the ideal part of the chemical potential 
with binding energy $E_c^{(0)}$ of isolated clusters ($\rm H,~H_2,~H^{-},~H_2^{-}$). 
The non-ideal parts $\Delta\mu_{cd}$ are defined by the second virial coefficients $b_{cd}$, similar further terms $\Delta\mu_{cde}$ etc. of Eq.(\ref{eq:mu}). 
The virial expansion is diverging for Coulomb interactions (for $e-e$, $i-i$, $e-i$ contributions).  
Therefore we consider screening interaction and take as an approximation for protons the classical Debye shift $\Delta_i=-\kappa e^2/2$ 
with the inverse screening length $\kappa^2=(\frac{4\pi n_i e^2}{k_B T})$. 
For electrons we use the Pad\'e formulae \cite{KSK}, which can be used at any plasma degeneracy 
\begin{eqnarray}
\Delta_e=\frac{\mu_{\rm D}-\frac{1}{2}(\pi\beta)^{-1/2}\bar{n}+8\bar{n}^2\mu_{\rm GB}}{1+8\ln[1+\frac{1}{16\sqrt{2}}(\pi\beta)^{1/4}\bar{n}^{1/2}]+8\bar{n}^2},
\label{eq:s151} 
\end{eqnarray}
where $\bar{n}=n_e\Lambda_e^3$, $\mu_{\rm D}=-(\pi\beta)^{-1/4}\bar{n}^{1/2}$ is the chemical potential in low-density limit (Debye limiting law), 
and  $\mu_{\rm GB}=-\frac{1.2217}{r_s}-0.08883\ln[1+\frac{6.2208}{r_s^{0.7}}]$ is the Gell-Mann-Brueckner approximation for the highly degenerate region 
(in $[\rm Ryd]$), 
$r_s=(3/(4\pi n_e))^{1/3}/a_{\rm B}$ is the Brueckner parameter.  

Finally, using the equations (\ref{eq:s1}) and (\ref{eq:mu})
the system of equations can be solved to derive the composition of the plasma 
with components $e, i, \rm H, \rm H_2, \rm H^-, \rm H_2^-$  
\begin{equation}
\begin{aligned}
\frac{n_{\rm H}}{n_e n_i}&=\Lambda_e^3 \exp(-\beta E_0)\exp\Big[\beta(\Delta_e+\Delta_i
\\
&-\Delta\mu_{\rm HH}^{\rm sc}+\Delta\mu_{e\rm H}^{\rm sc}+\Delta\mu_{e\rm H_2}^{\rm sc})\Big], 
 \\
 \frac{n_{\rm H_2}}{n_{\rm H}^2}&= b_{\rm H\rm H}^{\rm bound}\exp\Big[\beta (2\Delta\mu_{\rm HH}^{\rm sc}-\Delta\mu_{\rm H_2H_2}^{\rm sc})\Big],
 \\
 \frac{n_{\rm H^-}}{n_e n_{\rm H}} &=b_{{\rm H}e}^{\rm bound}\exp\Big[\beta(\Delta_e+\Delta\mu_{\rm HH}^{\rm sc})\Big],
 \\
 \frac{n_{\rm H^-_2}}{n_e n_{\rm H_2}} &=b_{{\rm H_2}e}^{\rm bound}\exp\Big[\beta(\Delta_e+\Delta\mu_{\rm H_2H_2}^{\rm sc})\Big],
 \\
 n_e^{\rm tot}&=n_e+n_{\rm H}+2n_{\rm H_2}+2n_{\rm H^-}+3n_{\rm H_2^-},
\end{aligned}
\label{eq:s3}
\end{equation} 
where 
$E_0=13.6$ eV is the ground state energy of hydrogen.  
The bound parts of the second virial coefficients are $b_{\rm HH}^{\rm bound}$, $b_{\rm He}^{\rm bound}$ and $b_{\rm H_2e}^{\rm bound}$. 
The dissociation energy of hydrogen molecule $D_0=4.75$ eV, the vibrational constant $h\nu/k_{\rm B}=6338.2 $ K \cite{HH}  and the rotational constant $B=87.58 $ K \cite{HH} 
are included in the bound part of the second virial coefficient 
\begin{equation}
b_{\rm HH}^{\rm bound}=\frac{1}{\sqrt{2}}\Lambda_{\rm H}^3 (\frac{T}{B})\frac{1}{1-\exp(-h\nu/k_BT)}\exp(\beta D_0). 
\label{eq:s13}
\end{equation}
$\Delta\mu_{cd}^{\rm sc}=-2n_{d} b_{cd}^{\rm sc}/\beta$ is scattering part of the non-ideal part of chemical potential for species $c,~d$. 
The second virial coefficient for $\rm H_2-H_2$ interaction
is treated using hard-core model $b_{\rm H_2\rm H_2}=\frac{2\pi}{3}d_{\rm H_2}^3(T)$. The diameter of hydrogen molecule $d_{\rm H_2}$ and  $b_{\rm HH}^{\rm sc}$
are taken from Ref. \cite{KSK}.  
Data for the second virial coefficient $b_{{\rm H}e}$ are taken from the previous work \cite{OFRR}. Other second virial 
coefficient $b_{{\rm H_2}e}$ are taken from our calculation, see Table \ref{tab:3}. 

\begin{figure}[htp]
\includegraphics[width=0.45\textwidth]{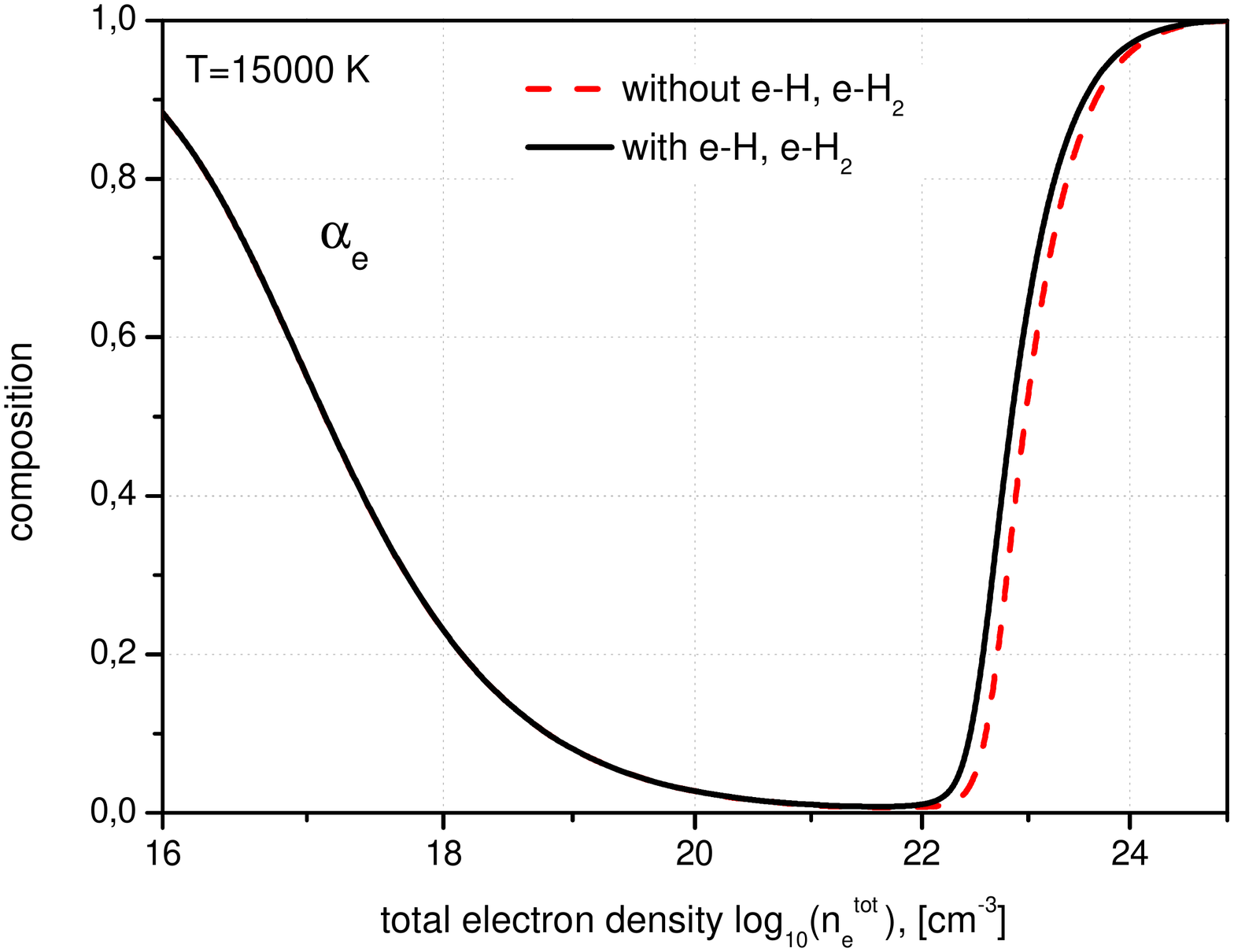}
\caption{(Color online) Fraction $\alpha_e$ with and without interaction terms at $T=15000$ K}
\label{fig:s4}  
\end{figure}

\begin{figure*}[htp]
\includegraphics[width=0.65\textwidth]{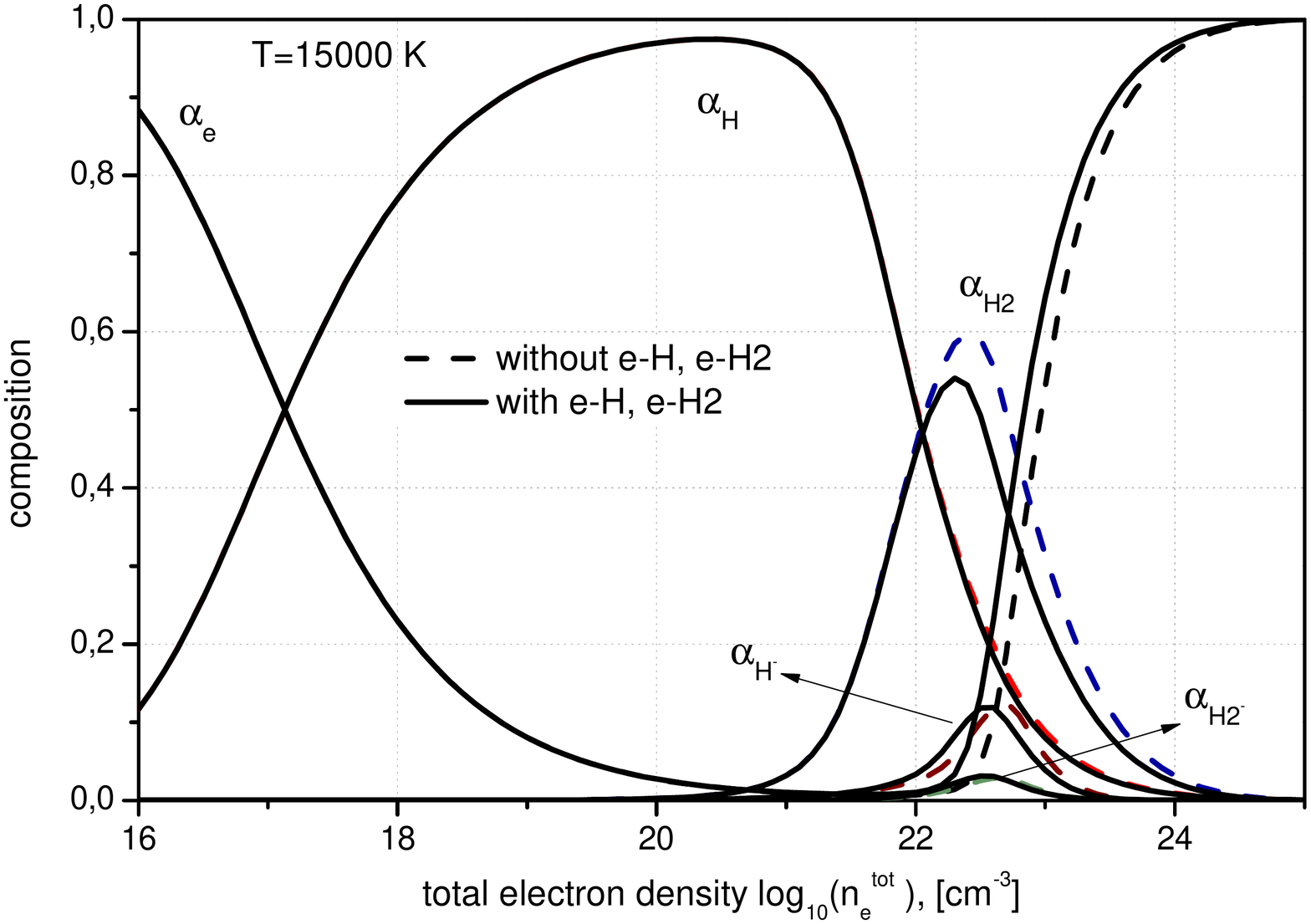}
\caption{(Color online) Composition for $e$, H, H$_2$, H$^-$, H$_2^-$ with and without interaction terms at $T=15000$ K}
\label{fig:s5}  
\end{figure*}

We focus on the influence of these two interactions ($e-\rm H$ and
$e-\rm H_2$) on the ionization equilibrium.  
Solution of the coupled equations (\ref{eq:s3}) for temperature $T=15000$ K are shown in the Figs. \ref{fig:s4} and \ref{fig:s5} in terms of fractions
$\alpha_e=n_e/n_e^{\rm tot}$,  $\alpha_{\rm H}=n_{\rm H}/n_e^{\rm tot}$, $\alpha_{\rm H_2}=2n_{\rm H_2}/n_e^{\rm tot}$, 
$\alpha_{\rm H^-}=2n_{\rm H^-}/n_e^{\rm tot}$ and 
$\alpha_{\rm H_2^-}=3n_{\rm H_2^-}/n_e^{\rm tot}$. Selected data are also given in Tables \ref{tab:s2} and \ref{tab:s3}.
Figure \ref{fig:s4} shows the results for the electron fraction with and without interaction 
terms $\Delta\mu_{e{\rm H}}^{\rm sc}$ and $\Delta\mu_{e{\rm H_2}}^{\rm sc}$.
Corrections are observed only at higher densities. 
 The inclusion of these additional interactions leads to an increase of the electron fraction at the same total electron density. 
Figure \ref{fig:s5} shows the comparison of all fractions with and without the interaction terms. 
Corrections due to to the additional interaction considered here can be seen at 
total electron densities $10^{21}-10^{22}$ cm$^{-3}$. At this density range, the fractions $\alpha_{\rm H}$, $\alpha_{\rm H_2}$ and $\alpha_e$ are relatively large and 
the interaction between electrons and atoms as well as molecules gives a significant contribution to the composition.  
\begin{table}[htp]
\caption{\label{tab:s2}Fractions for $e$, H, H$_2$, H$^-$ and H$_2^-$ with interaction terms for $e-\rm H$ and $e-\rm H_2$ at $T=15000$ K}
\centering
\begin{ruledtabular}
\begin{tabular}{c|c|c|c|c|c}
  $n_e^{\rm tot}$, [cm$^{-3}$] & $\alpha_e$& $\alpha_{\rm H}$&$\alpha_{\rm H_2}$&$\alpha_{\rm H^-}$&$\alpha_{\rm H_2^-}$ \\
\hline
10$^{16}$&0.883&0.117&1.09$\cdot10^{-8}$&4.15$\cdot10^{-7}$&1.32$\cdot10^{-13}$ \\ 
10$^{18}$&0.230&0.770&4.75$\cdot10^{-5}$& 6.94$\cdot10^{-5}$&1.46$\cdot10^{-8}$ \\
10$^{20}$&0.027&0.964&0.0077&0.000957&2.62$\cdot10^{-5}$\\
10$^{21}$&0.011&0.891&0.0939&0.00315&0.0011\\
10$^{22}$&0.039&0.381&0.531&0.0202&0.0292\\
10$^{23}$&0.816&0.050&0.129&0.00396&0.000204 \\   
\end{tabular} 
\end{ruledtabular}
\end{table}
\begin{table}[htp]
\caption{\label{tab:s3}Fractions for $e$, H, H$_2$, H$^-$ and H$_2^-$ without interaction terms for $e-\rm H$ and $e-\rm H_2$ at $T=15000$ K}
\centering
\begin{ruledtabular}
\begin{tabular}{c|c|c|c|c|c}
  $n_e^{\rm tot}$, [cm$^{-3}$] & $\alpha_e$& $\alpha_{\rm H}$&$\alpha_{\rm H_2}$&$\alpha_{\rm H^-}$&$\alpha_{\rm H_2^-}$ \\
\hline
10$^{16}$&0.883&0.117&1.09$\cdot10^{-8}$&4.15$\cdot10^{-7}$&1.32$\cdot10^{-13}$\\
10$^{18}$&0.230&0.770&4.75$\cdot10^{-5}$&6.94$\cdot10^{-5}$&1.45$\cdot10^{-8}$\\
10$^{20}$&0.027&0.964&0.0077&0.000967&2.59$\cdot10^{-5}$\\
10$^{21}$&0.009&0.892&0.094&0.003&0.001\\
10$^{22}$&0.007&0.387&0.566&0.014&0.025\\
10$^{23}$&0.534&0.069&0.322&0.031&0.043\\
\end{tabular} 
\end{ruledtabular}
\end{table}

\subsection{Comparison with the excluded volume approach}
The excluded volume concept is one of the popular simpler approximations to take the interaction of electrons with neutrals into account \cite{EFFGP}. 
The fraction of volume occupied by atoms can be defined with the filling parameter $\eta=4/3 \pi r_{\rm H}^3 n_{\rm H}$, where $r_{\rm H}$ is 
an atomic radius. The second virial coefficient is given for a system of hard spheres as  
\begin{equation}
 b^{\rm ex}_{e\rm H}=-\frac{2}{3} \pi r_{\rm H}^3. 
\label{eq:s26}
\end{equation}
The composition of partially ionized hydrogen plasma is calculated 
replacing $b^{\rm sc}_{{\rm H}e}$ in the previous calculations by the second virial coefficient from excluded volume concept 
$b^{\rm ex}_{e\rm H}$  
with the hard-core radius of $r_{\rm H}=1.0~a_{\rm B}$. Within the considered accuracy, this leads approximately to identical results as for the calculations without electron-atom interaction, see Table \ref{tab:s3}. 
It indicates that the excluded volume concept with $r_{\rm H}=1.0~a_{\rm B}$ makes the interaction of electrons and clusters negligible.

On the other hand, it is interesting to fit the radii of hydrogen atoms and molecules on the basis of the second virial coefficients according to Eq.(\ref{eq:s26}). 
Using the data for $b_{\rm He}$ from the previous paper Ref. \cite{OFRR} and for $b_{\rm H_2e}$ from Table \ref{tab:3}, one can obtain the corresponding radii. 
Table \ref{tab:s5} shows the results of the fit for different temperatures. The increase of temperature leads to smaller radii. High energy of projectile electrons leads to 
fast collisions and closer distances. In the last two columns of the Table \ref{tab:s5}, data from Ref. \cite{KSK} are given. 
The radii are obtained by fitting of the classical virial coefficients assuming real potential for 
$\rm H-H$ and $\rm H_2-H_2$ interactions. 

Note that the mean particle distance $d_e=d_i=(3/4\pi n_e^{\rm tot})^{1/3}$ gives a general limit of applicability of the cluster virial expansion for temperature 
and density parameters. At the total electron density $n_e^{\rm tot}=1.37 \cdot 10^{22} \rm cm^{-3} $ the mean particle distance is $d_e=4.90 ~a_{\rm B}$. That means, 
at $T=15000$ K the use of the second virial coefficients is possible up to this density. 

\begin{table*}[htp]
\caption{Hard core radius of hydrogen atom and molecule}
\centering
\begin{ruledtabular}
\begin{tabular}{cccccc}
$T$, [K] &$r_{\rm H}/a_{\rm B}$, full
&$r_{\rm H_2}/a_{\rm B}$, full,&$r_{\rm H}/a_{\rm B}$, Ref.\cite{KSK}& $r_{\rm H_2}/a_{\rm B}$, Ref.\cite{KSK}\\
\hline
10000&6.88&8.65&-&-\\
11000&6.33&8.25&-&-\\
12000&5.88&7.90&-&-\\
13000&5.50&7.58&-&-\\
14000&5.18&7.30&-&-\\
15000&4.90&7.05&1.50&1.78\\
20000&3.91&6.12&1.42&1.70\\
30000&2.91&5.03&1.30&1.57\\
50000&2.08&3.96&-&-\\
60000&1.86&3.65&-&-\\
70000&1.71&3.41&-&-\\
80000&1.59&3.21&-&-\\
90000&1.50&3.05&-&-\\
100000&1.43&2.91&-&-\\
   \end{tabular} 
\label{tab:s5}
\end{ruledtabular}
\end{table*}

\section{Conclusions\label{sec:Con}}
The cluster virial expansion for thermodynamical functions is considered for a partially ionized plasma. 
Using the Beth-Uhlenbeck formula \cite{BU}, the
second virial coefficient in the electron-molecule ($e-\rm H_2$) channel is related to phase shifts and possible
bound states in that channel. Results for
the $e -\rm H_2$ channel  are based on the polarization pseudopotential
model (\ref{eq:V2}) which is adapted to experimental data of scattering cross sections.  
A new bound state $\rm H_2^{-}$ occurs  in the bound part of the Beth-Uhlenbeck formula. Alternatively, 
it can be considered as a new constituent within the chemical picture. Our approach replaces the empirical hard-core model 
for the $e-\rm H_2$ by a more fundamental quantum statistical treatment. The influence on the equation of state and the composition is small and is
concentrated to that region where we have simultaneously the formation of $\rm H_2$ molecules as well as a large amount of free electrons.  


\section*{Acknowledgement}

This work is supported by the Deutsche Forschungsgesellschaft (DFG) under grant SFB 652 
``Strong Correlations and Collective Effects in Radiation Fields: Coulomb Systems, Clusters and Particles''.

\bibliography{literature}
\end{document}